\documentclass[conference]{IEEEtran}
\IEEEoverridecommandlockouts
\usepackage{cite}
\usepackage{enumitem,amsmath,amssymb,amsfonts}
\usepackage{algorithmic}
\usepackage{graphicx}
\usepackage{textcomp}
\usepackage{xcolor}
\usepackage[hyphens]{url}
\usepackage{booktabs}
\usepackage{multirow}
\usepackage{multicol}
\usepackage{fancyhdr,lipsum}
\usepackage[T1]{fontenc}
\usepackage{tabularx}
\usepackage{makecell}
\usepackage{upquote}
\usepackage{balance}
\usepackage[most]{tcolorbox}

\newlist{todolist}{itemize}{2}
\setlist[todolist]{label=$\square$}

\begin{document}

\title{Automatic Pull Request Title Generation}

\author{\IEEEauthorblockN{Ting Zhang, Ivana Clairine Irsan, Ferdian Thung, DongGyun Han, David Lo and Lingxiao Jiang}
\IEEEauthorblockA{School of Computing and Information Systems, Singapore Management University\\
Email:
\{tingzhang.2019,\:ivanairsan,\:ferdianthung,\:dhan,\:davidlo,\:lxjiang\}@smu.edu.sg
}}

\maketitle

\begin{abstract}
Pull Requests (PRs) are a mechanism on modern collaborative coding platforms, such as GitHub. 
PRs allow developers to tell others that their code changes are available for merging into another branch in a repository. A PR needs to be reviewed and approved by the core team of the repository before the changes are merged into the branch.
Usually, reviewers need to identify a PR that is in line with their interests before providing a review.
By default, PRs are arranged in a list view that shows the titles of PRs. Therefore, it is desirable to have a precise and concise title, which is beneficial for both reviewers and other developers. 
However, it is often the case that developers do not provide good titles; we find that many existing PR titles are either inappropriate in length (i.e., too short or too long) or fail to convey useful information, which may result in PR being ignored or rejected. Therefore, there is a need for automatic techniques to help developers draft high-quality titles.

In this paper, we introduce the task of automatic generation of PR titles. We formulate the task as a one-sentence summarization task. To facilitate the research on this task, we construct a dataset that consists of 43,816 PRs from 495 GitHub repositories. We evaluated the state-of-the-art summarization approaches for the automatic PR title generation task. We leverage ROUGE metrics to automatically evaluate the summarization approaches and conduct a manual evaluation. The experimental results indicate that BART is the best technique for generating satisfactory PR titles with ROUGE-1, ROUGE-2, and ROUGE-L F1-scores of 47.22, 25.27, and 43.12, respectively.
The manual evaluation also shows that the titles generated by BART are preferred.
\end{abstract}


\section{Introduction}
\label{sec:introduction}

As an emerging paradigm, pull-based software development has been widely applied in distributed software development~\cite{gousios2014exploratory,7180096}. 
With the pull-based development model, the main repository of the project is not shared for direct modification among \textit{contributors}~\cite{gousios2015work}.
Instead, contributors fork or clone the repository and make changes on their own branch.
When their changes are ready to be merged into the main branch, they create Pull Requests (PRs); the \textit{core team} (i.e., the \textit{integrators)} then review the changes made in the PRs to make sure that the changes satisfy functional (e.g., no compile error and test failure) and non-functional requirements (e.g., abide to coding convention). 
They may propose corrections, engage in discussion with the contributors, and eventually accept or reject the changes~\cite{gousios2016work}.
PRs are utilized in nearly half of the collaborative projects in GitHub~\cite{gousios2015work}.
Pull-request-based development is usually associated with reduced review times and a high number of contributions~\cite{zhu2016effectiveness}.

Generally, a PR consists of a title, a description (optional), and a set of commits. 
The PR title and description are designed to help the \textit{readers} (not limited to integrators, but refers to anyone reading the PR) grasp the context and purpose of the PR.
Frequently, a PR is linked with one or more issue reports.
Issue reports in issue tracking systems (e.g., GitHub issues) are used to keep track of bugs, enhancements, or other requests.
Therefore, many PRs contain the identifiers of the linked issues in their titles or descriptions.
Since contributors often neglect to write a PR description, the role of the PR titles is significant.
For example, 219,909 PRs (16.4\%) do not have descriptions among our collected 1,341,790 PRs.

Another common case is that PRs are displayed in a list view by default so that only the title and other meta-information (e.g., author name, tags, and PR id) are available.
In the absence of a PR description, a high-quality title becomes more important for readers to understand the intention of a PR.
There are other ways to figure out what a PR is about, such as direct interaction with the PR's owner, or checking the details like commit messages and linked issues.
However, these are certainly inefficient for software maintenance.

In addition, PRs usually have a fast turnaround: they are either processed fast or left open for a long time without merging to the main branch~\cite{gousios2014exploratory}.
In large projects, it is a challenge for integrators to handle a high volume of incoming PRs~\cite{gousios2015work}.
Prior research~\cite{guo2010characterizing} on bug reports has shown that well-written bug reports are more likely to gain the triager’s attention and influence on deciding whether a bug gets fixed.
Similarly, the quality of PR, such as length of title and description, has a significant impact on PR evaluation latency~\cite{yu2015wait}.
Intuitively, PR titles serve as the first criterion for integrators to decide whether certain PRs are relevant to their expertise, which can potentially speed up the review process.
Nowadays, the quality of PR descriptions has gained more attention in practice: many Git service providers support software maintainers to use templates to improve the quality of PR descriptions.\footnote{\url{https://docs.github.com/en/communities/using-templates-to-encourage-useful-issues-and-pull-requests/about-issue-and-pull-request-templates}}
However, there is no emphasis on helping developers compose high-quality PR titles.

To fill in this gap, we aim for automatic generation of PR titles to compose accurate and succinct PR titles.
We formulate the automatic PR title generation task as a one-sentence summarization task.
Its goal is to produce a concise and informative target sequence of tokens (\textit{title}), which is a summary of the given \textit{source} sequence of tokens.

As no prior work has been devoted to this task before, we conducted a comprehensive evaluation on the effectiveness of five state-of-the-art summarization methods on the automatic PR title generation, including both general-purpose and domain-specific summarization methods. 
Some of these methods are extractive (i.e., they extract the sentences from the source sequence and then concatenate them to form the target sequence), while others are abstractive (i.e., they generate the target sequence with sentences that are different from the original sentences in the source sequence).
For \textit{general-purpose} summarization methods, we have identified three approaches: BERTSumExt~\cite{liu2019text}, BART~\cite{lewis2019bart}, and Text-To-Text Transfer Transformer (T5) ~\cite{raffel2019exploring}.
BERTSumExt~\cite{liu2019text} is an extractive summarization method that utilizes the pre-trained bidirectional encoder representations from Transformers (BERT)~\cite{devlin2018bert}.
BART~\cite{lewis2019bart} and T5~\cite{raffel2019exploring} are two large pre-trained Transformer-based architectures, which can be used for abstractive summarization.
For \textit{domain-specific} summarization methods, we evaluate two methods, i.e., PRSummarizer~\cite{liu2019automatic} originally designed for PR description generation and iTAPE~\cite{chen2020stay} initially designed for issue title generation.
Another related paper is by Liu et al.~\cite{liu2022sotitle} that proposed a Stack Overflow title generation approach; However the underlying model of their method is T5, which is already included in the general-purpose summarization method.

Specifically, in the work, we would like to answer two Research Questions (RQs) to understand the performance of different methods on the PR title generation task: 

\textbf{RQ1:} \textit{In terms of automatic evaluation, how do different methods perform on the PR title generation task?}

\textbf{RQ2:} \textit{To what extent can the best performing approaches automatically generate PR titles as developers would do?}

Due to the lack of a suitable dataset, we construct a dataset named \textsc{PRTiger} (\underline{P}ull \underline{R}equest \underline{Ti}tle \underline{G}en\underline{er}ation), which is the first dataset that can be leveraged for PR title generation.
Our focus is to help contributors to compose non-trivial PR titles, which aims to help integrators to grasp the main changes made in the PRs.
We identified and applied several rules to keep the PR titles that suit the use scenario.
In the end, we have 43,816 PR titles belonging to 495 GitHub repositories in PRTiger.
The source sequence in the dataset is the concatenation of PR description, commit messages, and linked issue title, with an average length of 114 words.
The \textit{target} sequence is the corresponding PR title, with an average length of 7 words.

ROUGE metrics~\cite{lin2004rouge}, the standard metrics for automatic evaluation of summarization technique effectiveness, are adopted to evaluate model performance.
We also conducted a manual evaluation by inviting three evaluators.
For each sample, the evaluators were asked to score three titles by reading the source sequence.
The titles generated by the automatic methods and the original human-written titles were randomly shuffled.
Evaluation criteria include \textit{correctness}, \textit{naturalness}, and \textit{comprehensibility}. 
The details can be found in Section~\ref{sec:evaluation_metrics}.
The results suggest that BART outperforms the other techniques in terms of both automatic and manual evaluation.
To conclude, our main contributions are threefold:
\begin{itemize}[nosep,leftmargin=1em]
    \item{We introduce the task of automatic PR title generation.}
    \item{We construct a dataset named \textsc{PRTiger}, which consists of 43,816 PRs from GitHub for the PR title generation task. PRTiger is the first benchmark for this task.}
    \item{We conduct evaluation of state-of-the-art summarization methods, including extractive (Oracle extraction, BERTSumExt~\cite{liu2019text}) and abstractive ones (PRSummarizer~\cite{liu2019automatic}, iTAPE~\cite{chen2020stay}, BART~\cite{lewis2019bart}, and T5~\cite{raffel2019exploring}).
    The evaluation includes both automatic evaluation and manual evaluation from multiple aspects.
    The results show that BART outperforms other approaches by significant margins.}
\end{itemize}

The remainder of this paper is organized as follows: Section~\ref{sec:background} describes the background, including motivation, the usage scenario, and problem formulation.
Section~\ref{sec:dataset} gives details about the process to construct the PRTiger dataset.
We give an introduction to all the summarization methods we evaluated in Section~\ref{sec:methods}.
We describe the experimental settings in Section~\ref{sec:experiment} and present results of our experiments in Section~\ref{sec:result}. 
Section~\ref{sec:discussion} presents qualitative analysis and threats to validity.
We also discuss related work in Section~\ref{sec:related}.
We conclude this paper and present future directions in Section~\ref{sec:conclusion}.
\section{Background}
\label{sec:background}
In this section, we first present the motivation and the usage scenario of automatic PR title generation.
Next, we discuss about problem formulation and list some challenges.

\begin{table*}[t]
\caption{PR title examples}
\centering
\begin{tabularx}{\textwidth}[t]{|p{3.5cm}|X|p{4cm}|}
\hline

\textit{\textbf{Example 1}}  &  \textit{\textbf{Example 2}}  & \textit{\textbf{Example 3}} \\
\hline
 
\textbf{Title:} \textit{Various.} 
 
\textbf{Source:} \url{https://github.com/angular/angular/pull/49} 
  
\textbf{Description:} None

\textbf{Commit Message:}
\begin{itemize}
\item{\textit{feat(DartWriter): support string interpolation}}
\item{\textit{feat(facade/lang): support int}}
\item{\textit{refactor(lexer): rename to scanner, use ints, etc.}}
\end{itemize}
\textbf{Issue:} None
& 
\textbf{Title:} \textit{Please let me know if this is a sufficient translation and I can do more... I'm kinda new to rebase, so forgive me if I'm doing this wrong}
 
\textbf{Source:} \url{https://github.com/ant-design/ant-design/pull/6370}

\textbf{Description} \textit{First of all, thanks for your contribution! :-) 
Please makes sure these boxes are checked before submitting your PR, thank you!
\begin{todolist}
    \item Make sure you propose PR to correct branch: bugfix for master, feature for latest active branch feature-x.x.
    \item Make sure you follow antd's code convention.
    \item Run npm run lint and fix those errors before submitting in order to keep consistent code style.
    \item Rebase before creating a PR to keep commit history clear.
    \item Add some descriptions and refer relative issues for you PR.
\end{todolist}
}

\textbf{Commit Message:}
\begin{itemize}
    \item{\textit{testing translation}}
    \item{\textit{adding advanced search}}
\end{itemize}
\textbf{Issue:} None
& 
\textbf{Title:} \textit{Revert "Add godot version in backtrace message"}

\textbf{Source:} \url{https://github.com/godotengine/godot/pull/30221}

\textbf{Description:}  \textit{Reverts \#28572}

\textit{The idea is good, but we can't use the configurable crash handler message to display the version string, it should be hardcoded in the crash handler print code for each platform instead.
}

\textbf{Commit Message:}

\textit{Revert "Add godot version in backtrace message"}

\textbf{Issue:} None \\
\hline
\end{tabularx}
\label{tab:example}
\end{table*}

\subsection{Motivation}
A good PR title should accurately summarize what the PR is about.
An appropriate PR title would help the PR readers to set their minds on what things they would expect to read from the PR.
However, many PR titles failed to accurately and succinctly summarize the PR content.

For example, \textbf{Example 1} in Table~\ref{tab:example} has a very short title, with only one word.
It does not summarize the PR well.
In this example, the PR description is absent.
A good PR title would save readers' time from checking the details of commits.
The current title fails to make a clear summary of the PR.
Generally speaking, PR titles with very few words are unlikely to convey enough information for readers to grasp the PR content.

On the other hand, as admitted by the PR author that they had limited experience in rebase, \textbf{Example 2} in Table~\ref{tab:example} shows a relatively long title which failed to concisely summarize the PR as well.
In this example, the PR description is taken straight from a static template, which includes several checkboxes.
Yet, except for the template itself, the PR author did not write anything else in the description.
The current PR title itself is lengthy and does not summarize the changes.
Usually, lengthy titles do not succinctly summarize the source sequence and may be written by contributors with limited experience, as shown in Example 2.

These poorly-written PR titles motivate us to resort to an automatic approach to help contributors compose titles.
Automatic PR title generation methods would make the process of submitting new PRs and integrating PRs more efficient.
They are helpful to contributors as they can reduce contributors' burden on writing.
The generated high-quality PR title is then beneficial to integrators as well, as it can potentially save their time and speed up the reviewing process.

\subsection{Usage Scenario}
The considered usage scenario in our work could be stated as follows:
When a contributor is drafting a new PR, an automatic approach can help them compose a title that contains an overview of the changes made in the PR.
Ideally, the PR title should cover the information from commit messages and also the issue information if the PR is resolving or is linked to any issue.
This automatic title generation approach will save contributors' time from thinking about how to write a succinct summary for their PR contents, so they can focus on writing a detailed description.

\subsection{Problem Formulation}
We formulate the PR title generation task as a one-sentence summarization task~\cite{chen2020stay}.
For simplicity, we refer to a concatenation of PR descriptions, commit messages, and linked/resolved issue titles as a \textit{source} sequence, and the PR title as a \textit{target} sequence in the rest of our paper.
We will also use \textit{target} and \textit{title} to refer the PR title interchangeably.
For a PR title generation model, the input is the source sequence, and it is supposed to generate a PR title, which is the one-sentence summary of the source sequence.

In this work, we focus on generating non-bot written and non-\textit{trivial} PR titles.
\textit{Trivial} PR titles are those that are authored by humans and can be written with minimal effort since PRs handle common tasks.
One such example is \textbf{Example 3} in Table~\ref{tab:example}.
In this example, the PR description contains the ID and the description of PR that needs to be reverted.
The title \textit{Revert ``Add godot version in backtrace message"} is sufficient to understand the purpose of this PR. 
The type of PR titles does not ask for summarization ability.
A concatenation of a keyword (i.e., \texttt{revert}) and the title of the PR that needs to be reverted would suffice and considered to be a common and good practice.
For bot-written titles, contributors do not need to bother writing titles.
Thus, for bot-written and trivial titles, there is no need for an automatic approach to help compose PR titles.

The PR title generation task is similar to PR description generation~\cite{liu2019automatic} and title generation for issues~\cite{chen2020stay} and Stack Overflow posts~\cite{liu2022sotitle}.
However, the PR title generation task has its own unique challenges: 
(1) Compared to the title generation task for issues/Stack Overflow posts, in the PR title generation task, there is a semantic gap between different resources, i.e., PR description, commit messages, and issues. 
In particular, the two resources of our input sequence: the linked issue and the PR description are generally written by different authors (issue reporter and issue triager, respectively); 
(2) Compared to the PR description generation task, the target sentence (PR title) is quite short and it has to be as concise and precise as possible.   
The four tasks are complementary to each other and help reduce the workload of developers in different ways.
\section{Benchmarking Dataset Building}
\label{sec:dataset}

As there is no existing dataset for PR title generation, we decide to build the first benchmark dataset for this task. 
In this work, we experimented with GitHub data due to its popularity.
Although Liu et al.~\cite{liu2019automatic} shared a dataset for the PR description generation task, they did not include the titles of PRs in their dataset.
In addition, their dataset only contains engineered Java projects, which may lack diversity. 
Therefore, a new dataset containing PR titles gathered from various programming language repositories is needed. 
We first collected the data; next, we filtered out PRs which did not belong to the usage scenario we focused on in this work.
Then, we filtered out the content from PRs which do not help to generate accurate and succinct PR titles.

\subsection{Data Collection} 
To collect PR data from GitHub, we firstly got 7 repository lists: Top-100 most-starred and Top-100 most-forked repositories (regardless of programming language); Top-100 most-starred repositories that are written primarily in one of the following languages: JavaScript, Python, Java, C, and C++.\footnote{The detailed repository lists are available here: \url{https://github.com/EvanLi/Github-Ranking/blob/f4cf5694eaed7ad1ee59425d7c0dcf9f3e8511f9/Top100}}
The number of stars and the number of forks are two different metrics in GitHub:
The number of star means how many people click on the repository to show their appreciation.
A fork is a copy of a repository and the number of forks indicates how many people copied this project.
The Top-100 most-starred and most-forked repositories cover a wide range of programming languages, e.g., Shell (\texttt{ohmyzsh/ohmyzsh}), Go (\texttt{kubernetes/kubernetes}), and TypeScript (\texttt{microsoft/vscode}).
Given the 7 lists have common repositories, after removing redundancies, in total we crawled 578 distinct repositories.
We collected the PRs from each repository using GitHub GraphQL API.\footnote{\url{https://docs.github.com/en/graphql}}
For each repository, we only kept the merged PRs that were published up to December 31, 2021.
Given a merged PR published before the year 2022, we retrieved its title, description, commit messages, and linked issues.
In total, we collected 1,341,790 merged PRs from the 578 GitHub repositories.

\subsection{Data Preprocessing}

\begin{table}[t]
\caption{Our trivial PR title patterns, and example PR titles from the collected dataset.}
\centering
\begin{tabularx}{0.48\textwidth}{|l|X|}
\hline
\textbf{Starts with (lowercased)} & \textbf{Example Title} \\
\hline
automated cherry pick of & \textit{Automated cherry pick of \#12022 \#13148 upstream release 1.0}  \\
\hline
merge to & \textit{Merge to live part 2 on 4-27}\\
\hline
revert & \textit{Revert "Add log\_only to debug messages"} \\
\hline
rolling up & \textit{Rolling up changes to staging from master} \\ 
\hline
rolling down & \textit{Rolling down changes from staging to master} \\
\hline
roll engine & \textit{Roll engine dart roll 20180920} \\
\hline
rollup of & \textit{Rollup of 5 pull requests} \\
\hline
roll plugins & \textit{Roll Plugins from 6d8ea78c5da1 to 361567b9189c (4 revisions)} \\ 
\hline
update live with current master & \textit{Update live with current master} \\
\hline
\end{tabularx}
\label{tab:template}
\end{table}

\textbf{Selecting PRs.}
For each PR, to better simulate the scenario when a contributor opens a new PR, we first removed the commits submitted after the PR was created.
Then, following Liu et al.~\cite{liu2019automatic}, we filtered out the PRs which have less than two commits or more than 20 commits.
As Liu et al. pointed out, we can directly use the commit message as the PR title if a PR only contains one commit, and a PR with too many commits is usually used for synchronization purpose instead of being a typical contribution from developers.
We also removed the PRs (1) containing non-ASCII characters; (2) authored by bots.
In addition, we also filtered out the PRs which contain \textit{trivial} titles, where automatic methods for generating PR titles are not needed.
We mainly identified the following four types of trivial titles:
(1) \textit{Recurrent titles.} Table~\ref{tab:template} shows the templates which occurred prominently in our collected PRs.
If any PR title starts with these patterns, we exclude it.
(2) \textit{Too short or too long titles.} Following the dataset building rules used by iTAPE~\cite{chen2020stay}, we excluded the titles with less than 5 words or more than 15 words.
Similar to Chen et al.~\cite{chen2020stay}, we observe that PR titles having 5-15 words are of reasonably appropriate length to precisely and succinctly describe key ideas.
(3) \textit{Titles with limited overlap with the source sequence.} If 20\% of the words in the title were not present in the source sequence, we considered the title not to be a good summary of the source sequence and therefore excluded it from the dataset.
(4) \textit{Titles were copied from the source sequence.} We first lowercased both the title and the source sequence.
If the title can be exactly matched to the description or concatenation of the commit messages, we removed the PR from the dataset as we consider this PR as a bad example to train the model.

\begin{table}[t]
\caption{Data statistics on our collected pull requests}
\centering
\begin{tabular}{|r|r|r|r|}
\hline
\textbf{With < 2 or > 20} & \textbf{With non-ASCII} &  \textbf{Authored} & \textbf{With trivial} \\
\textbf{commits} & \textbf{characters} & \textbf{by bot} & \textbf{titles} \\
\hline
 1,147,734 & 16,396  & 1,642 &  111,780 \\
\hline
\hline
\multicolumn{2}{|c|}{{\bf PRs Left}} & \multicolumn{2}{c|}{{\bf Total PRs Collected}} \\
\hline
\multicolumn{2}{|r|}{51,753} & \multicolumn{2}{r|}{1,341,790} \\
\hline
\end{tabular}
\label{tab:all_data}
\end{table}

After selecting PRs based on the above criteria, we have 51,753 PRs left in the dataset.
Table~\ref{tab:all_data} shows the statistics of our collected PRs.

\textbf{Cleaning the selected PRs.} We followed iTAPE~\cite{chen2020stay} to remove tags in the PR titles. We also followed PRSummarizer~\cite{liu2019automatic} to (1) remove checklists in the source sequence; and (2) remove identifiers in both source and target sequence.
We also added extra steps: (1) Removing PR templates in the source sequence. 
We first queried through the GitHub API to find out whether a repository provided a PR template for contributors to compose a PR description.
If there was a PR template, we saved the template string.
In our dataset, 214 out of 495 repositories provided a PR template at the time we called the GitHub API.
To remove the template information from each PR, we first split the PR description into lines. 
Then, for each line in the source sequence, if it can be matched exactly to the template string, we removed it.
Otherwise, we kept the lines.
By doing this, we reduced the noise in the source sequence. 
(2) Removing automatically generated commit messages in the source sequence.
We observed that many commit messages only convey the \textit{merge branch} information, e.g., \textit{Merge branch `master' into tst-indexing-16}.
We used the following four regular expressions to remove the automatically generated commit messages: (1) \texttt{merge .*? branch .*? into} (2) \texttt{merge branch .*? into} (3) \texttt{merge pull request \textbackslash \#} and (4) \texttt{merge branch \textbackslash \textquotesingle}.

\begin{table}[t]
\caption{Data statistics on different splits}
\centering
\begin{tabular}{@{}l llllll@{}}
\toprule
& \multicolumn{2}{c}{\textbf{Train}}  & 
\multicolumn{2}{c}{\textbf{Validation}} &
\multicolumn{2}{c}{\textbf{Test}} \\
\cmidrule(lr){2-3}\cmidrule(lr){4-5}\cmidrule(lr){6-7}
& \textbf{Source} & \textbf{Target} & \textbf{Source} & \textbf{Target} & \textbf{Source} & \textbf{Target} \\
\midrule

\textbf{Instance \#} & 35,052 & 35,052 & 4,382 & 4,382 & 4,382 & 4,382\\
\midrule
\textbf{Avg Word \#} & 114 & 7 & 114 & 7 & 112 & 7 \\
\bottomrule
\end{tabular}
\label{tab:dataset}
\end{table}

After applying the pre-processing steps to our data, we further excluded PRs which have less than 30 words or more than 1,000 words.
In the end, we have a dataset \textit{PRTiger} consisting of 43,816 PRs for experiments.
We split PRTiger into training, validation, and test sets with a ratio of 8:1:1.
Table~\ref{tab:dataset} shows the number of instances and the average word count in the train, validation, and test set respectively.
\section{Summarization Methods}
\label{sec:methods}
In this section, we elaborate on the summarization methods evaluated in this work.
Summarization methods can be broadly categorized into two groups, i.e., \textit{extractive methods} and \textit{abstractive methods}~\cite{allahyari2017text}.
An \textit{extractive} summarization method extracts sentences from the source sequence and then concatenates them to form the target sequence. 
In contrast, an \textit{abstractive} summarization method represents the source sequence in an intermediate representation. It then generates the target sequence with sentences that are different from the original sentences in the source sequence~\cite{el2021automatic}.
We experimented with state-of-the-art extractive and abstractive methods.
Their details can be found as follows:

\subsection{Extractive Methods} 
We experimented with two extractive summarization methods, i.e., oracle extraction and BertSumExt~\cite{liu2019text}.

\textbf{Oracle Extraction}: This is not a \textit{real} approach.
Instead, it can be viewed as an upper bound of extractive summarization methods or serve as a measure to gauge the capability of other extractive summarization methods.
Oracle extraction scores have been commonly used in summarization literature for comparison purpose~\cite{cachola2020tldr,yuan2021can}.
It may have different variants depends on the specific task setting.
In our work, oracle extraction selects the sentence from the source sequence that generates the highest ROUGE-2 F1-score compared to the original title.
We firstly split the PR description, commit messages, and issue titles into sentences.
Next, we computed the ROUGE-2 F1-score of each sentence with the original PR title.
We selected the sentence with the highest ROUGE-2 F1-score as the generated title by this method. 
Since oracle extraction needs the original title as a reference, it cannot be applied in practice.
However, it can be used for comparison to understand other extractive methods' performance.
    
\textbf{BertSumExt~\cite{liu2019text}}: BertSumExt is built on top of BERT-based encoder by stacking several inter-sentence Transformer layers.
Specifically, for each sentence in the source sequence, BertSumExt represents each sentence with the vector of the \textit{i}-th \texttt{[CLS]} symbol from the top layer.
Then, several inter-sentence Transformer layers are then stacked on top of BERT outputs, to capture document-level features for extracting sentences as the summary.
BertSumExt achieves better results with less requirement on the model, compared to other models that enhance the summarization via the copy mechanism~\cite{gu2016incorporating}, reinforcement learning~\cite{paulus2017deep}, and multiple communicating encoders~\cite{celikyilmaz-etal-2018-deep}.
Identical to what we did in oracle extraction, we also split PR description, commit messages, and issue titles into sentences.
BertSumExt will then give a score to each sentence based on the suitability of becoming a summary. 
BertSumExt was initially evaluated on three single-document news summarization datasets, which could be summarized in a few sentences. 
In the original implementation, BertSumExt chooses the sentences with Top-3 highest scores as a summary.
As the PR title generation task was formulated as a one-sentence summarization task, we take the Top-1 sentence as the generated PR title.

\subsection{Abstractive Methods} 
The two most-similar works are identified, i.e., PRSummarizer~\cite{liu2019automatic} and iTAPE~\cite{chen2020stay}. 
As they did not directly choose the sentence from the source sequence, we also categorized them into abstractive methods. 
Besides, we applied BART~\cite{lewis2019bart} and T5~\cite{raffel2019exploring}, which are state-of-the-art method for text summarization. (we put the exact model version of the Hugging Face \texttt{transformers} library that we use \footnote{\url{https://huggingface.co/models}} in parentheses):

\textbf{PRSummarizer~\cite{liu2019automatic}}: This text summarization model was designed to automatically generate PR descriptions from the commits submitted with the corresponding PRs, which is a sequence-to-sequence (Seq2seq) learning task.
The underlying model of PRSummarizer is the attentional encoder-decoder model~\cite{bahdanau2014neural}.
Besides, PRSummarizer can handle two unique challenges, i.e., out-of-vocabulary (OOV) words and the gap between the training loss function of Seq2seq models and the discrete evaluation metric ROUGE.
Especially, PRSummarizer copes with OOV words by integrating the pointer generator~\cite{see2017get}.
The pointer generator either selects a token from the fixed vocabulary or it will copy one token from the source sequence at each decoding step.
To minimize the gap between the loss function and the ROUGE metrics, PRSummarizer also leverages a reinforcement learning (RL) technique named self-critical sequence training~\cite{rennie2017self} and adopts a particular loss function named RL loss~\cite{paulus2017deep}.

\textbf{iTAPE~\cite{chen2020stay}}: iTAPE uses a Seq2seq based model to generate the issue title using the issue body. 
Specifically, iTAPE adopts the attentional RNN encoder-decoder model.
In addition, to help the model process the low-frequency human-named tokens effectively, iTAPE firstly inserts additional ``tag tokens'' before and after each human-named token, i.e., identifiers and version numbers. 
These tag tokens are added to the issue body to indicate their latent semantic meanings.
Furthermore, iTAPE adopts a copy mechanism~\cite{gu2016incorporating}, which allows the model to copy tokens from the input sequence.
The underlying architecture is the pointer-generator~\cite{see2017get}, a commonly used abstractive summarization approach before the dominant usage of pre-trained models.

\textbf{BART~\cite{lewis2019bart}} (\texttt{facebook/bart-base}): is a Seq2seq autoencoder based on a standard Transformer~\cite{vaswani2017attention} architecture.
The pre-training process of BART consists of two stages: (1) corrupt the input text by using an arbitrary noising function and (2) a Seq2seq model is learned to reconstruct the original text by minimizing the cross-entropy loss between the decoder output and the original sequence.
A number of noising approaches are evaluated in the BART paper.
The best performance is achieved by adopting both the noising methods, i.e., (1) randomly shuffling the order of the original sentences, and (2) applying an in-filling scheme, where a single mask token is used to replace arbitrary length spans of text (including zero length).
BART was pre-trained with the same data as RoBERTa~\cite{liu2019roberta}, i.e., 160GB of news, books, stories, and web text. 
BART achieves new state-of-the-art results on several text generation tasks, including abstractive dialogue, question answering, and summarization tasks.

\textbf{T5~\cite{raffel2019exploring}} (\texttt{t5-small}): T5 is a pre-trained language model which aims to convert all NLP tasks into a unified text-to-text-format where the input and output are always text strings.
The advantage of this T5 text-to-text framework is that we can use the same model, loss function, and hyper-parameters on any NLP task.
T5 is also based on the Transformer architecture.
Similar to BART, T5 was pre-trained on a masked language modeling objective: contiguous spans of token in the input sequence are replaced with a mask token and the model is trained to reconstruct the masked-out tokens.
Unlike BART, T5 was pre-trained with the Colossal Clean Crawled Corpus (C4) dataset, which consists of 750GB of English text from the public Common Crawl web scrape.
T5 was reported to achieve state-of-the-art results on many benchmarks including summarization, question answering, and text classification.
\section{Experimental Settings}
\label{sec:experiment}
This section describes the relevant design and settings of our study.
We list two research questions
and the evaluation metrics, and briefly describe the implementation details.

\subsection{Research Questions}
We would like to empirically evaluate the performance of different approaches on the automatic PR title generation task. 
The study aims to answer the following RQs:

\textbf{RQ1:} \textit{In terms of automatic evaluation, how do different methods perform on the PR title generation task?}
Although there is no prior work on automatically generating PR titles, we choose the approaches from the two closest works~\cite{liu2019automatic,chen2020stay} as the baselines. 
We used the implementations of these two approaches on our task. 
For comparison, we also evaluated the state-of-the-art general-purpose extractive and abstractive summarization techniques.

\textbf{RQ2:} \textit{To what extent can the best performing approaches automatically generate PR titles as developers would do?}
Other than providing the results on automatic evaluation, we also conducted a manual evaluation. 
Given the expense to run manual evaluation, we only evaluated the two best-performing methods on the automatic evaluation.
We invited three annotators who are not an author of this paper.

\subsection{Evaluation Metrics}
\label{sec:evaluation_metrics}
\textbf{Automatic Evaluation}
Following the prior works~\cite{liu2019automatic,chen2020stay}, we use Recall-Oriented Understudy for Gisting Evaluation (ROUGE)~\cite{lin2004rouge} to measure the quality of generated summaries for the summarization task.
ROUGE-N measures the overlap of n-grams~\cite{lin2004rouge} between the model-generated summary and the reference summary.
The formulas to calculate ROUGE-N can be shown as follows:
\begin{equation*}
    R_{ROUGE-N} = \frac{Count(overlapped\_N\_grams)}{Count(N\_grams\_in\_reference\_summary)}
\end{equation*}
\begin{equation*}
    P_{ROUGE-N} = \frac{Count(overlapped\_N\_grams)}{Count(N\_grams\_in\_generated\_summary)}
\end{equation*}
\begin{equation*}
    F1_{ROUGE-N} = 2\times \frac{R_{ROUGE-N}\times P_{ROUGE-N}}{R_{ROUGE-N}+ P_{ROUGE-N}}
\end{equation*}
As the variable names suggest, R (recall) measures the percentage of the N-grams in the reference summary that has been covered by the generated summary, and P (precision) presents the percentage of N-grams in the generated summary that is, in fact, relevant or needed.
F1-score of the ROUGE scores is used to represent and give an equal importance between recall and precision.
In this task, we report the precision, recall, and F1-score of ROUGE-N (N=1,2) and ROUGE-L from each method. 
ROUGE-1, ROUGE-2, and ROUGE-L are commonly used in the literature to understand the summary quality~\cite{liu2019automatic,chen2020stay}.
ROUGE-1 and ROUGE-2 measure the overlap of uni-grams (1-grams) and bi-grams (2-grams), respectively. 
A uni-gram consists of a single word, while a bi-gram consists of two consecutive words.
Instead of n-grams, ROUGE-L measures the longest common subsequence between the reference summary and the generated summary.
We treat F1-score of ROUGE as the main summarization performance measure.
We first get the generated summaries from each approach.
For the ROUGE scores calculation, we adopt the metric implemented in Hugging Face \texttt{datasets} library~\cite{lhoest-etal-2021-datasets}.

\begin{table}[t]
\caption{An example PR with the original title and the titles generated by BART and T5. https://github.com/grpc/grpc/pull/1655}
\centering
\begin{tabularx}{0.48\textwidth}{|l|X|}
\hline
\textbf{Source} & <desc> this fixes \#1551. servers now respond to unparseable arguments with the invalid\_argument status and the text of the deserialization error, instead of crashing. </desc> <cmt> added failing tests for server bad argument handling </cmt> <cmt> fixed server to handle invalid arguments without breaking </cmt> <iss> node rpc server cannot recover from malformed requests </iss>
\\
\hline
\textbf{Original Title} & handle invalid arguments sent to the node server \\
\hline
\textbf{BART} & fix server bad argument handling \\
\hline
\textbf{T5} & fix server to handle invalid arguments without breaking node \\
\hline
\end{tabularx}
\label{tab:qualitative-2}
\end{table}

\textbf{Manual Evaluation}
In addition to automatic evaluation, we also conducted a manual evaluation to better understand and evaluate the quality of titles generated by different approaches.
Since ROUGE scores are calculated based on the overlap of n-grams, the generated summaries may be semantically incorrect, even with very high ROUGE scores. 
This is due to the ROUGE scores limitation which only measures the lexical similarity between two sentences. 
Hence, it cannot gauge the comprehensibility of the summaries generated by the model.~\cite{VANDERLEE2021101151}
Thus, we sampled 150 PRs from the test set. 
With this sample size, we can maintain the confidence level of 92\% with an 8\% margin of error.
Three evaluators were invited to give the quality scores to the PR titles generated by the two best approaches and the original titles written by developers. 
All of the evaluators have more than 6-year experience in programming and more than 5-year experience using GitHub.
They have a basic understanding of how pull-based software development works (i.e., they are experienced in making changes, pushing commits, writing issue reports, and opening PRs).

The techniques used to generate titles are hidden from the evaluators; they cannot judge based on the bias of knowing the authorship.
For each sample, evaluators were provided with the \textit{source} sequence along with three \textit{titles}: two of the titles are generated by the two best-performing approaches, i.e., BART and T5, and the rest is the original title.
We randomly shuffled the order of the three titles.
To help the evaluators read the source sequence clearly, we use \texttt{<desc></desc>}, \texttt{<cmt></cmt>}, and \texttt{<iss></iss>} to enclose description, commit messages, and linked issue titles, respectively.
One sample source sequence can be seen in Table~\ref{tab:qualitative-2}.
Evaluators were required to read through the \textit{source} sequence and the three titles.
They were asked to score the three titles (1 - very poor; 5 - very good) with regards to the following aspects:
\begin{itemize}[nosep,leftmargin=1em]
    \item{\textbf{Correctness:} To which extent, do you think the title \textit{correctly} summarize the source sequence?}
    \item{\textbf{Naturalness:} To which extent, do you think the title is resembling a human-written title?}
    \item{\textbf{Comprehensibility:} To which extent, do you think the title is easy to \textit{understand}?}
\end{itemize}
Additionally, they were also required to rank the three titles.
Their personal preference for these titles does not necessarily base on the three criteria listed above.
If the titles are the same, they can rank two titles with the same rank.
Otherwise, they must give different ranks for each title.

\subsection{Implementation Details} 
We run all the experiments with NVIDIA Tesla V100 GPUs.
For iTAPE, we preprocessed the identifier and version numbers with the scripts provided by the authors.\footnote{\url{https://github.com/imcsq/iTAPE}}
For PRSummarizer, we changed the vocabulary size from 50k to 200k to handle the OOV issue. 
Given BART (\texttt{BART-base}, which contains 140 millon parameters\footnote{\url{https://github.com/pytorch/fairseq/blob/main/examples/bart/README.md#pre-trained-models}}) has more parameters than T5 (\texttt{T5-small}, which contains 60 million parameters\footnote{\url{https://github.com/google-research/text-to-text-transfer-transformer#released-model-checkpoints}}), we run BART with the batch size of 4; while 8 for T5.
All the remaining hyper-parameters were left as the default values.
The detailed default values can be found in our replication package.\footnote{\url{https://github.com/soarsmu/PRTiger}}
\section{Experimental Results}
\label{sec:result}

\begin{table*}[t]
\caption{Automatic evaluation Results on the test dataset}
\centering
\begin{tabular}{@{}lcrrrrrrrrr@{}}
\toprule
\multirow{2}{*}{\textbf{Approach}} &  \multirow{2}{*}{\textbf{Avg. Length}} &
\multicolumn{3}{c}{\textbf{ROUGE-1}}  & 
\multicolumn{3}{c}{\textbf{ROUGE-2}} &
\multicolumn{3}{c}{\textbf{ROUGE-L}}  \\
\cmidrule(lr){3-5}\cmidrule(lr){6-8}\cmidrule(lr){9-11}
 & & \textbf{Precision} & \textbf{Recall} & \textbf{F1} & \textbf{Precision} & \textbf{Recall} & \textbf{F1} & \textbf{Precision} & \textbf{Recall} & \textbf{F1}  \\
\midrule
\textit{Extractive} &&& \\
\;--\;\textbf{Oracle Extraction} & 13 & 47.61 & 55.74 & 46.97 & 32.51 & 34.04 & 30.25 & 44.88 & 51.33 & 43.91  \\
\;--\;\textbf{BertSumExt} & 13 &36.18& 44.95 & 37.64 & 18.18 & 21.71 & 18.48 & 32.76 & 40.28 & 33.94 \\
\midrule
\textit{Abstractive} &&& \\
\;--\;\textbf{PRSummarizer} & 6 & 41.4 & 36.68 & 37.91 & 20.06 & 17.26 & 17.99 & 38.22 & 33.83 & 34.98 \\
\;--\;\textbf{iTAPE} & 8 & 31.63 & 34.62 & 32.23 & 12.67 & 13.98 & 12.91 &  28.71 & 31.51 & 29.31 \\
\;--\;\textbf{BART} & 7 & 50.03 & 46.98 & 47.22 &  27.15 & 24.96 & 25.27 &45.71 & 42.85 & 43.12 \\
\;--\;\textbf{T5} & 7 & 44.88 & 42.15 & 42.06 & 23.22 & 21.3 & 21.39 &  41.09 & 38.49 & 38.46 \\
\bottomrule
\end{tabular}
\label{tab:result}
\end{table*}

\begin{table}[t]
\caption{Performance comparison with other summarization tasks: issue title generation (Chen et al. [13]), PR description generation (Liu et al. [12]), and lay language summarization of biomedical scientific reviews (Guo et al. [30])}
\centering
\begin{tabular}{@{}|l|c|c|c|@{}}
\hline
\multirow{2}{*}{\textbf{Method}} & \textbf{ROUGE-1} & \textbf{ROUGE-2} & \textbf{ROUGE-L}  \\
& \textbf{F1} & \textbf{F1} & \textbf{F1} \\
\hline
\textbf{BART in our work} & 47.22 & 25.27 & 43.12 \\ 
\hline

\textbf{Chen et al.~\cite{chen2020stay}} & 31.36 & 13.12 & 27.79 \\
\textbf{Liu et al.~\cite{liu2019automatic}} & 34.15 & 22.38 & 32.41 \\
\textbf{Guo et al.~\cite{guo2020automated}} & 53.02 & 22.06 & 50.24 \\
\hline
\end{tabular}
\label{tab:cross-result}
\end{table}

\subsection{RQ1: Comparison on Automatic Evaluation}
To investigate how different approaches perform on the PRTiger dataset, we firstly analyze model performance with ROUGE metrics.
Table~\ref{tab:result} shows the ROUGE scores from the six approaches.

Firstly, looking at the length of generated titles, we can observe that the extractive approaches produce longer titles than the abstractive approaches.
The average length of the titles generated by extractive approaches highly relies on the length of sentences selected from the dataset.
Although we set the extractive approaches to select a single sentence as a title, they select longer titles.
All the abstractive methods generate titles in the average length of 6-8 words.
The length range is similar to the average length across all the data splits (See Table~\ref{tab:dataset}).
It indicates that abstractive methods generate titles with an appropriate length, as it is capable of generating titles with the average length similar to the average length of the original titles.

Serving as the upper bound of extractive summarization approaches, Oracle Extraction gives the highest ROUGE-1, ROUGE-2, and ROUGE-L F1-scores since it relies on the original title.
In comparison, BertSumExt gave ROUGE-1, ROUGE-2, and ROUGE-L F1-score of 37.64, 18.48, and 33.94. 
These scores were 20\%, 38.9\%, and 22.7\% lower than the Oracle Extraction.
It suggests that, in practice, PR titles have a considerable amount of overlap with the source sequence.
Yet, BertSumExt is not capable of selecting the correct sentence every time.

Among the four abstractive approaches, BART and T5 achieved better performance than the two other abstractive approaches.
It demonstrates the power of pre-training in the PR title generation.
BART and T5 were both pre-trained with large general-purpose corpora.
Unlike these two approaches, PRSummarizer and iTAPE were trained solely on the PRTiger data, where PRSummarizer produces a better performance than iTAPE.
Considering PRSummarizer was originally proposed for PR description generation, the data they used to evaluate their approach has similar characteristics as PRTiger.
In addition, both PRSummarizer and iTAPE adopt the pointer generator~\cite{see2017get}.
It suggests the importance of RL-loss in PRSummarizer, which is used to minimize the gap between the loss function and the ROUGE metrics.

Comparing extractive and abstractive methods, BART shows an on-par performance as the Oracle Extraction, which indicates that BART is capable of capturing key points from the source sequence and generating precise PR titles.
BART outperforms the second-best approach T5 by 12.2\%, 18.1\%, and 12.1\%, in terms of ROUGE-1, ROUGE-2, and ROUGE-L F1-scores, respectively.
Although T5 gives worse performance than BART, it still outperforms the other approaches, i.e., PRSummarizer, by 10.9\%, 18.9\%, and 10\% with regards to ROUGE-1, ROUGE-2, and ROUGE-L F1-scores, respectively.
Interestingly, BertSumExt gives an on-par performance as PRSummarizer and it outperforms iTAPE.
Given the words in the PR, titles are not necessarily contained in the source sequence, the PR title generation task is naturally an abstractive summarization task.
However, as an extractive approach, BertSumExt shows relatively good performance.
It indicates that, in practice, developers may draft titles based on existing sentences, either in PR description or commit messages.

Other than comparing different approaches on the automatic PR title generation task solely, we show the ROUGE F1-scores from other related tasks.
Chen et al.~\cite{chen2020stay} work on the issue title generation task, where the proposed approach is named iTAPE and is evaluated in our work. 
Liu et al.~\cite{liu2019automatic} work on the PR description generation task, where the proposed approach is named PRSummarizer and is evaluated in our work as well.
Guo et al.~\cite{guo2020automated} work on automated generation of lay language summaries of biomedical scientiﬁc reviews.
The ROUGE F1-scores from the best performing method of these three tasks are present in Table~\ref{tab:cross-result}.
According to the result, we can find that BART in our work achieves a comparable-level of performance on the automatic PR title generation task.

\begin{tcolorbox}[colback=gray!5!white,colframe=gray!75!black,boxrule=0.2mm]
\textit{Automatic Evaluation} The fine-tuned BART and T5 outperform all the other approaches.
BertSumExt is on par with the two existing approaches (PRSummarizier and iTAPE).
The best-performing approach, BART, outperforms the second-best approach by 12.2\%, 18.1\%, and 12.1\%, in terms of ROUGE-1, ROUGE-2, and ROUGE-L F1-scores.
\end{tcolorbox}

\subsection{RQ2: Comparison on Manual Evaluation}

\begin{figure}[t]
    \centering
    \includegraphics[width=0.4\textwidth]{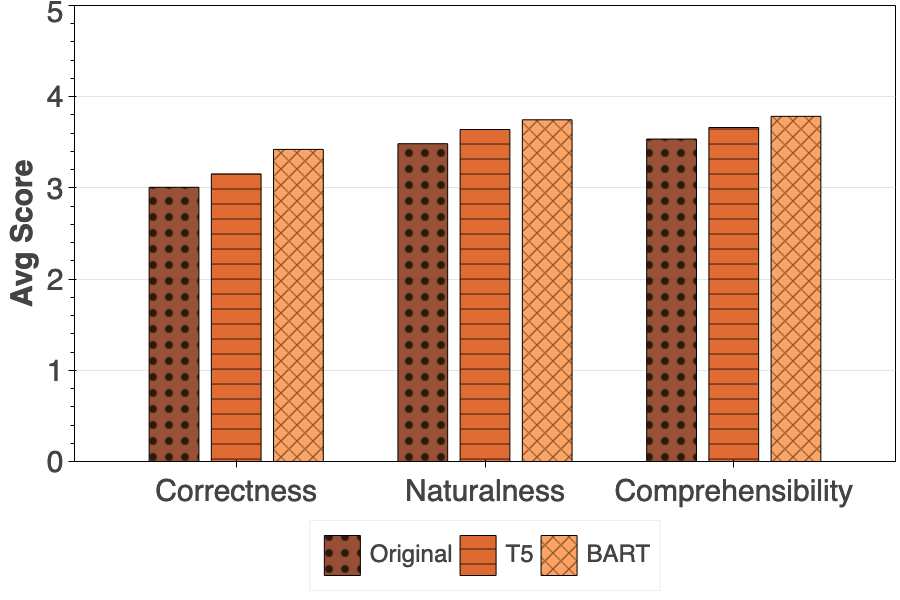}
    \caption{Average scores from three evaluators}
    \label{fig:score}
\end{figure}

\begin{figure}[t]
    \centering
    \includegraphics[width=0.4\textwidth]{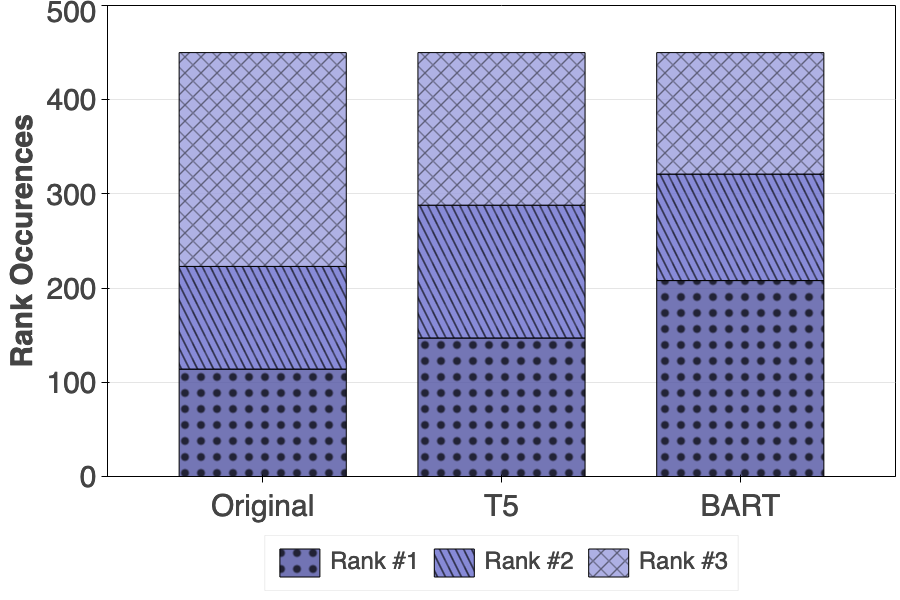}
    \caption{Rank occurrences for the original titles and the titles generated by BART and T5.}
    \label{fig:rank}
\end{figure}

Figure~\ref{fig:score} shows the average scores regarding three aspects from the three evaluators.
BART is better than T5 in the three aspects, which is in line with the automatic evaluation results with ROUGE metrics.
Surprisingly, we can see that BART and T5 show higher scores than the original titles in all three perspectives.
It indicates that the titles generated by automatic methods are more acknowledged than the titles written by PR authors.
Across the three aspects, we can find that all the three approaches receive the lowest average scores of \textit{correctness}.
They achieve slightly higher average scores regards \textit{comprehensibility} than \textit{naturalness}.
These results show that although PR titles read like written by humans and are easy to understand, they may not be correct enough from the evaluators' perspectives.

Figure~\ref{fig:rank} shows that for each rank, the percentage of different approaches.
Although the ranking is a relative order, it is allowed to be subjective and solely based on the evaluators' personal preference.
For the best ones, we can find that BART has been listed as the best one the most times.
T5 comes the second.
Again, surprisingly, the original title is less preferred.

While a larger-scale study is required, our work provides preliminary evidence that automatically generated PR titles are readable and comprehensible to PR readers, i.e., have higher scores in terms of correctness, naturalness, and comprehensibility.
In the sampled PR titles, automatically generated PR titles are preferred over the original titles in most cases.

\begin{tcolorbox}[colback=gray!5!white,colframe=gray!75!black,boxrule=0.2mm]
  \textit{Manual Evaluation} The PR titles generated by BART gained the highest scores on correctness, naturalness, and comprehensibility. T5 performs the second. They are both preferred more than the original titles.
\end{tcolorbox}

\section{Discussion}
\label{sec:discussion}
\subsection{Qualitative Analysis}

\begin{table}[t]
\caption{An example PR with the original title and the generated titles from five approaches: https://github.com/iluwatar/java-design-patterns/pull/1066}
\centering
\begin{tabularx}{0.48\textwidth}{|l|X|}
\hline
\textbf{Source Sequence} & reduces \textbf{checkstyle errors for} patterns: \textbf{api-gateway lazy-loading leader-election} changes involved java docs reordering imports indentations line length issues   reduces \textbf{checkstyle errors} in \textbf{lazy-loading}   reduces \textbf{checkstyle errors} in \textbf{leader-election}   reduces \textbf{checkstyle errors} in \textbf{api-gateway}
\\
\hline
\textbf{Original Title} & \textbf{resolves checkstyle errors for api-gateway, lazy-loading, leader-election} \\
\hline
\textbf{BART} & \textbf{resolves checkstyle errors for api-gateway lazy-loading leader-election} \\
\hline
\textbf{T5} & reduces \textbf{checkstyle errors for} patterns \\
\hline
\textbf{BertSumExt} & reduces  \textbf{checkstyle errors for} patterns : \textbf{api - gateway lazy - loading leader - election} changes involved java docs reordering imports indentations line length issues \\
\hline
\textbf{PRSummarizer} & reduces \textbf{checkstyle errors for} patterns: \textbf{api-gateway} \\
\hline
\textbf{iTAPE} & resolves \textbf{checkstyle errors for} patterns : \textbf{api-gateway lazy-loading} \\
\hline
\end{tabularx}
\label{tab:qualitative}
\end{table}

Composing PR titles is a non-trivial task, as it requires both summarization ability and domain knowledge.
Moreover, PR titles are generally short. 
Here, we would like to seek a qualitative understanding of the performance difference between different approaches.

\textbf{Automatic Evaluation.} In Table~\ref{tab:qualitative}, we showcase one example of titles produced by each approach and the original title along with the source sequence.
Note the source sequence is the cleaned version used for the models to generate titles.
We use boldface to highlight the common part between the text and the original title.
All the approaches can correctly generate \texttt{checkstyle errors for}, which occurs several times in the source sequence.
However, all the approaches except BART generate the first word as \texttt{reduces}, and they also generate \texttt{pattern}. 
It is natural, as the first sentence in the source sequence contains \texttt{reduces} and \texttt{pattern} as well.
It is interesting to see that BART does not directly take the first sentence.
Besides, the body does not even have the word \texttt{resolve}.
We checked the dataset and found that several PR titles in the same repository (\texttt{iluwater/java-design-pattern}) followed the same name style: \texttt{resolves checkstyle errors for}. 
The following text differs among PRs.
From this example, we can find that BART can learn this title style instead of simply choosing the first sentence.
BART does not only use exactly the same words from the source sequence.
Instead, it could correctly generate the words which are not present in the source sequence.
It shows BART has the promising ability to be adopted in the automatic PR title generation task.

\textbf{Manual Evaluation.} In Table~\ref{tab:qualitative-2}, we show an example from our sampled 150 PRs. 
All the evaluators indicate their preference among the three titles as: BART > T5 > Original.
Firstly, this PR contains description, two commit messages and linked issue.
The original title uses \texttt{handle}, which only appears in one of the commit messages.
Besides, the phrase \texttt{sent to} does not exist in the source sequence.
In comparison, the generated title from BART and T5 used \texttt{fix}, which occur twice in the source sequence and all the words in the generated titles are present in the source sequence.
The title generated by BART is shorter and summarizes the source sequence well, while the title generated by T5 is longer and covers more words from the source sequence.

We investigated PR titles with high comprehensibility scores (i.e., all evaluators gave a score of >=4 for comprehensibility). 
We found that a PR title is considered to be more comprehensible when: (1) It covers multiple sources of information, e.g., sentences in the PR description and commit messages. (2) It explicitly states that it fixes an issue or adds a feature (e.g., usually starts with the word "fixed" or "added"). 
We also investigated PR titles with low comprehensibility scores (i.e., all evaluators gave a score <=3 for comprehensibility). We found that a PR title is considered to be less comprehensible when: (1) the PR contains many commits but the PR title only takes information from a few commit messages. 
(2) The PR title has little connection with the PR description.

\subsection{Threats to Validity}
\textbf{Threats to internal validity} relate to the experimental bias.
Following the prior works in summarization studies~\cite{liu2019automatic,chen2020stay}, we adopted both automatic evaluation (i.e., ROUGE metrics) and manual evaluation. 
Like other manual-involved evaluations, our experimental results may be biased or inconsistent.
To mitigate this issue, we recruited 3 evaluators with 6+ years of computer programming and 5+ years of experience in using GitHub.
They are familiar with the issue and code review mechanism in GitHub.

Moreover, during the dataset building process, although we have performed the selection heuristics, the remaining PR titles that we use as ground truth for automatic evaluation may not be the ideal ones. 
We use them as proxies of the ideal reference titles for scalability reasons. 
There are a total of 43,816 titles in our experimental dataset and crafting the best possible reference titles for all of them manually is not practically feasible. 
We address this limitation of the automatic evaluation by manual evaluation; each of these evaluations has its own limitations: automatic (quality of reference titles) vs. manual (limited number of titles are considered).

\textbf{Threats to external validity} relate to whether our findings can be generalized to other datasets. 
To alleviate the external threats, we consider repositories from different programming languages diverse. 
We also considered two metrics in GitHub: Top-100 most-starred and Top-100 most-forked repositories.
Although we only included the repositories from GitHub, we do not identify a large difference between repositories from GitHub and those from other git service providers~\cite{bhattacharjee2020exploratory}.
Therefore, we consider the external threat to be minimal.
\section{Related Work}
\label{sec:related}
In this section, we review the two lines of research most related to our work: understanding pull requests and automatic software artifact generation.

\textbf{Understanding Pull Requests.}
Pull-based software development has attracted more and more interest in research. 
Some works focus on empirically understanding PRs.
Gousios et al. conducted two surveys to analyze the work practices and challenges in pull-based development from the integrator's~\cite{gousios2015work} and contributor's~\cite{gousios2016work} perspectives, respectively. 
In the first piece of work, they found that integrators successfully use PRs to solicit external contributions. 
Integrators concern with two major factors in their daily work, i.e., (1) quality: both at the source code level and tests (2) prioritization: typically they need to manage a large number of contribution requests at the same time. 
In the later one, they performed an exploratory investigation of contributors to projects hosted on GitHub. Their findings include, but are not limited to, the fact that contributors are very interested in knowing project status for inspiration and to avoid duplicating work, but they are not actively trying to publicize the PRs they are preparing.

Other works are interested in solving PR-related tasks.
Yu et at.~\cite{yu2016reviewer} studied the reviewer recommendation for PRs to improve the PR evaluation process.
They found that traditional approaches (e.g., SVM-based) for bug triage are also feasible for PR reviewer recommendations.
They also found that combining the social factors (e.g., common interests among developers) and technical factors (e.g., developer's expertise) is efficient to recommend reviewers.
Jiang et al.~\cite{jiang2021recommending} studied the tag recommendation for PRs.
They proposed FNNRec, which uses a feed-forward neural network to analyze titles, descriptions, file paths, and contributors, to help developers choose tags.
These prior works motivate our work to automatically generate high-quality PR titles to benefit both developers and reviewers.
Besides, good PR titles can also be helpful for downstream tasks which utilize PR titles.

\textbf{Automatic Software Artifact Generation.} 
Automatic software artifact generation has gained emerging interests in SE research.
Existing works range from automatically generating release notes~\cite{moreno2014automatic}, bug reports~\cite{li2018unsupervised}, to commit messages~\cite{xu2019commit}. 
Moreno et al.~\cite{moreno2014automatic} introduced an automatic approach to generate release notes. The proposed approach extracts changes from the source code, summarizes them, and integrates the summary of changes with information from versioning systems and issue trackers.
Li et al.~\cite{li2018unsupervised} proposed an unsupervised approach for bug report summarization. The approach leverages an auto-encoder network with evaluation enhancement and predefined fields enhancement modules.
Xu et al.~\cite{xu2019commit} proposed CODISUM to address the two limitations of prior commit message generation task, i.e., ignoring the code structure information and suffering from the OOV issue.
Specifically, to better learn the code changes representations, they first extract both code structure and code semantics from the source code changes, and then jointly model these two information sources.
Moreover, they adopted a copy mechanism to mitigate the OOV issue.
We also mention iTAPE, an issue title generation approach in Section~\ref{sec:methods}.
Our work and these works are complementary. 
We all aim to promote automatic generation in SE, while concentrating on different aspects.

\section{Conclusion and Future Work}
\label{sec:conclusion}
In this paper, we propose the task of automatic generation of PR titles. 
To facilitate the research on this task, we construct a dataset named PRTiger, which consists of 43,816 PRs from 495 GitHub repositories.
We conducted both automatic and manual evaluations on the state-of-the-art summarization approaches for the automatic PR title generation task.
The experimental results indicate that BART is the most capable technique for generating satisfactory PR titles with ROUGE-1, ROUGE-2, and ROUGE-L F1-scores of 47.22, 25.27, and 43.12, respectively.
Manual evaluation also shows that the titles generated by BART are preferred.

We believe that our work opens up many interesting research opportunities.
To name a few, one possible research direction is to consider the characteristics of PR data to propose a domain-specific pre-trained model. Domain-specific models are promising, such as BERTweet~\cite{nguyen2020bertweet}, which is pre-trained on English tweets, outperforms the strong baseline RoBERTa\textsubscript{base}~\cite{liu2019roberta} on three Tweet NLP tasks.
Additionally, to further improve the PR title generation performance, another direction is to leverage the hierarchical code structure information from code changes.

\balance

\vspace{0.2cm}\noindent {\bf Acknowledgment.}
This research / project is supported by the National Research Foundation, Singapore, under its Industry Alignment Fund – Pre-positioning (IAF-PP) Funding Initiative. Any opinions, findings and conclusions or recommendations expressed in this material are those of the author(s) and do not reflect the views of National Research Foundation, Singapore.

\bibliographystyle{IEEEtran}
\bibliography{main}

\end{document}